\documentclass[prf,superscriptaddress,a4paper,longbibliography,amsmath,amssymb,floatfix,10pt]{revtex4-1}
\bibstyle{apsrev}
\usepackage[utf8]{inputenc}
\usepackage[english]{babel}
\usepackage[section]{placeins}
\usepackage{bm}
\usepackage{xcolor}
\usepackage{float}
\usepackage{physics}
\usepackage{hyperref}
\usepackage{natbib}
\hypersetup{
    colorlinks=true,
    linkcolor=blue,
    citecolor=blue,
    filecolor=blue,
    urlcolor=blue 
}
\usepackage{cleveref}
\Crefname{equation}{Eq.}{Eqs.}
\Crefname{figure}{Fig.}{Figs.}
\Crefname{section}{Sec.}{Secs.}

\usepackage{siunitx}

\usepackage{tabularx}
\usepackage{placeins}

\usepackage{caption}
\usepackage{subcaption}
\usepackage{graphicx}
\usepackage{array}
\usepackage{soul}

\newcommand{\newtext}[1]{\textcolor{black}{#1}}

\begin{document}

%\title{Bubble-induced clogging and unclogging in porous transport layers modelled as constricted capillaries}
%\title{Bubbles at the Bottleneck: Clogging and Unclogging in Highly Porous Media Modeled as Constricted Capillaries}
%\title{When Bubbles Block the Flow: Clogging in Constrictions and Highly Porous Media}
%\title{Bubbles at the Bottleneck: Clogging in Constrictions and Highly Porous Media}

%\title{Bubble clogging, unclogging, and migration across porous transport layers}
%Bubble Clogging in Constrictions and Highly Porous Media
%Bubble Clogging in Constrictions and Porous Transport Layers}
%Bubble Clogging and Unclogging in Porous Transport Layers}
%\title{Bubbles at Constrictions: Clogging and Unclogging in Highly Porous Media}
\title{Bubbles in highly porous media: Clogging and unclogging at constrictions}
% Bubble-induced clogging and unclogging of constricted capillaries
% Bubble-induced clogging and unclogging in porous media modelled as constricted capillaries
% Bubble-induced clogging and unclogging in porous transport layers modelled as constricted capillaries

\author{J.M.P. Beunen}
\email{j.beunen@fz-juelich.de}
\affiliation{Helmholtz Institute Erlangen-Nürnberg für Erneuerbare Energien (IET--2), Cauerstr.~1, 91058 Erlangen, Germany}%
\affiliation{Department of Chemical and Biological Engineering, Friedrich-Alexander-Universit\"at Erlangen-N\"urnberg, Cauerstr.~1, 91058 Erlangen, Germany}

\author{T. Lappan}
\email{t.lappan@hzdr.de}
\affiliation{Institute of Fluid Dynamics, Helmholtz-Zentrum Dresden-Rossendorf, 01328 Dresden, Germany}%
\affiliation{Institute of Process Engineering and Environmental Technology, Technische Universität Dresden, 01062 Dresden, Germany}%

\author{P. Malgaretti}
\email{p.malgaretti@fz-juelich.de}
\affiliation{Helmholtz Institute Erlangen-Nürnberg für Erneuerbare Energien (IET--2), Cauerstr.~1, 91058 Erlangen, Germany}%

\author{O. Aouane}
\email{o.aouane@fz-juelich.de}
\affiliation{Helmholtz Institute Erlangen-Nürnberg für Erneuerbare Energien (IET--2), Cauerstr.~1, 91058 Erlangen, Germany}%

\author{K. Eckert}
\email{k.eckert@hzdr.de}
\affiliation{Institute of Fluid Dynamics, Helmholtz-Zentrum Dresden-Rossendorf, 01328 Dresden, Germany}%
\affiliation{Institute of Process Engineering and Environmental Technology, Technische Universität Dresden, 01062 Dresden, Germany}%

\author{J. Harting}
\email{j.harting@fz-juelich.de}
\affiliation{Helmholtz Institute Erlangen-Nürnberg für Erneuerbare Energien (IET--2), Cauerstr.~1, 91058 Erlangen, Germany}%
\affiliation{Department of Chemical and Biological Engineering and Department of Physics, Friedrich-Alexander-Universit\"at Erlangen-N\"urnberg, Cauerstr.~1, 91058 Erlangen, Germany}%

\begin{abstract}
Gas bubble transport through highly porous transport layers (PTLs) is a key
process in electrochemical devices such as proton exchange membrane water
electrolyzers, where bubbles generated at catalyst surfaces must migrate through
complex porous networks. To understand this process, we focus on model systems,
namely the motion of single, paired and multiple bubbles in capillaries and
study these by combining analytical modeling, three-dimensional color-gradient
lattice Boltzmann simulations, and X-ray radiography.
For single bubbles, we derive an analytical expression for the critical Bond number separating passage from clogging and show that, in the low deformation regime, it accurately predicts this transition in circular capillaries. Extending the study to bubble pairs, we uncover additional clogging and unclogging pathways, including hydrodynamic unclogging driven by pressure buildup in the interbubble film, and coalescence-induced clogging and unclogging. By mapping our results as functions of confinement ratio and Bond number, we define distinct dynamical regimes that control bubble passage.
Experiments on bubble chains rising through highly porous nickel foams confirm the predicted clogging and unclogging mechanisms.

\end{abstract}

\maketitle

\section{Introduction}\label{sec:introduction}

Bubble formation and transport in porous transport layers (PTLs) play a critical role in the performance of many electrochemical devices, including water electrolyzers~\cite{Carmo2013ComprehensiveReviewPEM,Pham2021PEMWE,Ivanova2023TechnologicalPathwaysProduce}, fuel cells~\cite{Bazylak2007EffectCompressionLiquid,Forner-Cuenca2015EngineeredWaterHighways}, and electrochemical reactors~\cite{Angulo2020InfluenceBubblesEnergy}. 
In proton exchange membrane water electrolyzers (PEMWEs), gas is continuously generated at catalyst surfaces and must be transported through a porous network before leaving the device~\cite{Carmo2013ComprehensiveReviewPEM,Suermann2015InvestigationMassTransport}. 
The porous transport layer acts as a multifunctional component that simultaneously provides water supply, gas removal pathways, electronic conduction, and thermal management within the cell~\cite{Bazylak2009LiquidWaterVisualization,Zinser2019AnalysisMassTransport}. 
The presence, motion, and accumulation of gas bubbles inside the PTL strongly influence mass transport, electrical conductivity, and pressure losses, and can ultimately limit device efficiency and stability~\cite{Suermann2015InvestigationMassTransport,Angulo2020InfluenceBubblesEnergy}.
Understanding the physical mechanisms that govern bubble motion through the complex pore geometry of PTLs is therefore essential for improving the design and operation of electrochemical systems.

At the pore scale, the transport of bubbles through PTLs is controlled by the interplay between buoyancy-driven forces, capillary resistance, and hydrodynamic stresses within the liquid phase. Because PTLs consist of interconnected pores with varying diameters, gas bubbles frequently encounter constrictions between adjacent pores that act as bottlenecks for transport. Whether a bubble deforms and passes through such a constriction or becomes trapped depends on the balance between the driving forces acting on the bubble and the capillary pressure required to enter the narrow opening. As a result, localized clogging events may occur, temporarily blocking pores and altering the effective permeability of the porous network.

In addition to the above-mentioned PTLs and catalytic systems~\cite{Vesztergom2021HydrogenBubbleTemplated,Solymosi2022NucleationRatedeterminingStep,Scheel2024EnhancementBubbleTransport}, bubble dynamics across porous materials also plays a critical role in microfluidic diagnostics~\cite{Huang2023MicrofluidicMethodsGeneration, Khanjani2025CapillaryMicrofluidicsDiagnostic}, and in geological flows where the migration of trapped gas influences permeability and phase distribution~\cite{Wang2021CapillaryEquilibriumBubbles}. In all of these systems, constrictions function as bottlenecks where capillary barriers arise, and flow may transition sharply between passage and clogging.
 
The fundamental physics of a bubble approaching a constriction is determined by the balance between the driving force and the interfacial tension, commonly expressed by the Bond number. If surface tension dominates, the bubble retains a nearly spherical shape and may pin at the constriction entrance. Once the forcing is sufficiently strong, the bubble deforms and enters the throat. This interplay between curvature, confinement, and forcing underlies classical descriptions of snap-off and capillary penetration, which have been studied in detail in experiments, theory, and simulations~\cite{Legait1983InertiaViscosityCapillary, Gauglitz1988ExperimentalDeterminationGasbubble, Ransohoff1987SnapoffGasBubbles, Marmur1988PenetrationSmallDrop,Tsai1994DynamicsDropConstricted, Jensen2004CloggingPressureBubbles,Gu2023NumericalStudyDroplet}.
Related analyses for droplets and soft particles squeezing through narrow openings~\cite{Zhang2017DropletSqueezingNarrow, Zhang2018ParticleSqueezingNarrow} formalize the same principle: geometric confinement amplifies the capillary barrier that must be overcome for entry.
Similar principles arise in the transport of deformable capsules and vesicles through narrow geometries. Barakat and Shaqfeh~\cite{Barakat2018SteadyMotionClosely}
showed that strong lubrication resistance and membrane tension buildup determine the pressure drop for steady particle motion in tight tubes. It was further demonstrated by simulations that confinement, membrane mechanics, and the thinning of the lubricating film govern whether an object elongates, slows down, or becomes arrested~\cite{Kaoui2012HowDoesConfinement,Kusters2014ForcedTransportDeformable,Fai2017ActiveElastohydrodynamicsVesicles}.

If bubbles always appeared in isolation, these classical mechanisms would fully describe clogging and passage in constricted geometries. However, for bubbles in succession, hydrodynamic coupling through the intervening film, lubrication-pressure buildup, and coalescence~\cite{Bashkatov2023H2BubbleMotion,Bashkatov2025ElectrolyteDropletSpraying} introduce alternative pathways for clogging or unclogging. Experiments and simulations on snap-off, film formation, coalescence, capsule suspensions, and emulsion flow~\cite{Yan2006NumericalStudyCoalescence,Pena2009SnapoffLiquidDrop, Cobos2009FlowOilWater, Roman2017MeasurementsSimulationLiquid,Hoang2018ThreedimensionalSimulationDroplet,Munir2023TwodimensionalNumericalModelling,Gu2023NumericalStudyDroplet,Bielinski2021SqueezingMultipleSoft} illustrate that slight variations in spacing, confinement, or wetting~\cite{Rox2025DualWettingElectrode} can strongly affect such confined multiphase flows.

Despite this broad foundation, the collective transport of successive bubbles through a constriction has to the best of our knowledge not been characterized systematically. A trailing bubble may accelerate or delay a leading bubble, increase the local pressure through lubrication, or trigger coalescence that either assists passage or produces a larger bubble that clogs the throat. These interactions depend on a nonlinear combination of geometrical confinement, Bond-number-controlled forcing, and interbubble spacing.

Following this, characterizing the dynamics of bubbles in a porous transport layer is a tremendous task, and it requires accounting for the heterogeneity of such a medium. Clearly, such complexity prevents us from an overall understanding of the physical mechanisms governing the dynamics of the bubbles. Accordingly, to capture the interplay of the different physical processes involved in the dynamics of bubbles across porous transport layers, we focus on a model porous system, namely, a circular channel with a single constriction.

We construct a dimensionless framework that quantifies confinement, the relative bubble size to the throat, and the hydrodynamic coupling between consecutive bubbles. This allows us to map the transitions between passage, clogging, breakup, and coalescence, and to identify the conditions under which a trailing bubble generates lubrication forces that promote unclogging. We examine these behaviors through numerical simulations and validate them with X-ray radiographic measurements of bubble chains rising through hydrophilic nickel foams, which provide a porous analogue with intrinsic geometrical variability. Together, these results establish how bubble interactions alter clogging behavior in confined flows and provide a quantitative basis for predicting when bubbles will pass or obstruct a constriction.

The remainder of this paper is organized as follows: \Cref{sec:problem-statement} introduces the geometry and the governing dimensionless numbers. \Cref{sec:analytical-model} explains the analytical model for bubble entry into a constriction and \Cref{sec:numerical-method} describes the numerical method. Our simulation results are presented in \Cref{sec:simulation-results} and accompanied by experimental results in \Cref{sec:experiments}. In \Cref{sec:conclusion}, we conclude.

\section{Problem statement}\label{sec:problem-statement}

\begin{figure}[!b]
    \centering
    \includegraphics{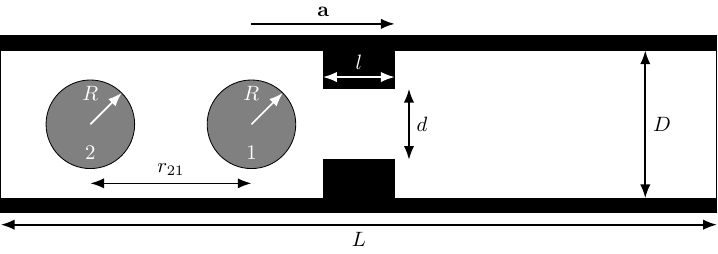}
    \caption{Schematic of a constricted circular capillary containing two bubbles. The channel has a diameter $D$ and a length $L$, with a constriction of diameter $d$ and length $l$. Bubbles 1 and 2 have radius $R$ and are separated by a center-to-center distance $r_{21}$. A gravity-induced acceleration $\mathbf{a}$ acts on both bubbles. Note that the dimensions are not drawn to scale.}
    \label{fig:schematic_problem_statement}
\end{figure}

We model a single constriction of a PTL as a cylindrical channel with diameter $D$ and length $L$ combined with an abrupt cylindrical constriction of diameter $d$ and length $l$. The constriction length $l$ is kept fixed, while its diameter $d$ varies. One or two bubbles of radius $R$ are placed in the channel. A two-dimensional schematic of the setup is shown in \Cref{fig:schematic_problem_statement}.

In all simulations, we label the leading bubble with index $1$. This bubble is initialized close to the constriction, which could be considered somewhat artificial. However, \newtext{for the problem being studied, viscous and inertial} effects \newtext{can be considered} negligible. \newtext{In all relevant cases, the capillary number
\begin{equation}
    \mathrm{Ca} = \frac{\rho_{\rm bubble} \nu v}{\sigma} = \frac{m \nu v}{V \sigma} 
    \label{eq:capillary_number}
\end{equation}
and the Weber number
\begin{equation}
    \mathrm{We} = \frac{\rho_{\rm bubble} v^2 d}{\sigma} = \frac{m v^2 d}{V \sigma}
    \label{eq:weber_number}
\end{equation}
are much smaller than one. Here, $\nu$ the bubble's kinematic viscosity, while $\rho_{\rm bubble}$ corresponds to the density inside the bubble. This allows to define the capillary and Weber numbers in terms of the bubble mass $m$ and the undeformed bubble volume $V$. The average velocity while passing the constriction is denoted by $v$, and the surface tension between the bubble and the carrier fluid is denoted by $\sigma$. As a result, buoyancy} and surface-tension forces \newtext{are} dominating, \newtext{and} the initial \newtext{momentum} of the bubble before entering the constriction is not important.

In simulations with two bubbles, the additional trailing bubble is labeled with index $2$, and the initial center-to-center separation is denoted by $r_{21}$. A gravity-induced acceleration $\mathbf{a}$ acts only on the bubbles, mimicking the effect of buoyancy. The solid walls are assumed to be strongly preferentially wetted by the carrier fluid, corresponding to an equilibrium contact angle of $\theta = 30^\circ$. With these variables defined, we can now specify several dimensionless numbers that characterize the transport of bubbles through the capillary.

First, we define two dimensionless numbers that quantify the capillary geometry. Given that we are mainly interested in the behavior of bubbles through capillaries in highly porous structures, we assume that the constriction is significantly shorter than the length required to contain the bubble fully. In other words, we assume that the constriction volume $V_{c} = \frac{1}{4} \pi d^2 l$ is significantly smaller than the bubble volume $V = \frac{4}{3} \pi R^3$ in all cases. To this end, we define the dimensionless length confinement ratio $\mathrm{C}_l = l / \left(2 R\right) = 0.2$, which ensures that the bubble is never fully inside the channel. \newtext{Given that the goal of the paper is to model the behavior of bubbles moving through highly porous media, we can reasonably assume the constriction length to be very small. To this end, we opt to keep the constriction length small and constant in the simulations, and we focus on the case in which the bubble does not fully fit into the constriction, $V_c < V$.} Furthermore, we define the dimensionless confinement ratio
\begin{equation}
    \mathrm{C} = \frac{d}{2 R},
    \label{eq:confinement_ratio}
\end{equation}
based on the ratio of the constriction diameter and the bubble radius. The confinement is left as a variable and determines the resistance to bubble passage. The smaller the confinement ratio, the greater the increase in Laplace pressure required for the bubble to enter the constriction, thereby increasing the work required for the bubble to pass.

Second, we define a dimensionless number for the initialization in a simulation with two bubbles. The center-to-center distance at initialization can be linked to the bubble radius with a bubble offset ratio
\begin{equation}
    \mathrm{B}_{\text{offset}} = \frac{r_{21}}{2 R}.
    \label{eq:bubble_offset_ratio}
\end{equation}
Here, a bubble offset ratio of one indicates that the bubbles have no carrier fluid between them and will coalesce immediately. Similarly, when the bubble offset ratio increases, the thickness of the lubrication layer increases as well. As a result, the amount of carrier fluid that needs to be squeezed out of the interbubble gap before possible coalescence also increases.

Finally, we define the Bond number, quantifying the ratio between the buoyancy force due to gravity and the capillary forces experienced by the bubble when crossing the channel as
\begin{equation}
    \mathrm{Bo} = \frac{m \left|\mathbf{a}\right| R^{2}}{V\sigma}.
    \label{eq:Bond_number}
\end{equation}
Here, $\left|\mathbf{a}\right|$ is the magnitude of the gravity-induced acceleration vector. The bubble radius $R$ is chosen as a characteristic length scale. Similarly, this Bond number can indicate the resistance to deformation as the bubble enters the constriction. \newtext{Furthermore, it can be straightforwardly mapped to a more common definition of the Bond number used in experiments, using a density difference between the carrier fluid and bubble $\Delta \rho$ and gravitational acceleration $g$.}

\newtext{In our theoretical work and simulations, we can set the carrier fluid and the bubble to have the same mass density. We can do this because surface tension forces dominate over viscous and inertial effects. Additionally, the density ratio between a carrier fluid and bubbles is generally very high in experiments. Therefore, the gravity-induced force in our definition of the Bond number, $m \left|\mathbf{a}\right|$, is equivalent to the force that would be exerted in experiments if the same volume of carrier fluid were displaced. In other words,
\begin{equation}
    \Delta \rho g \approx \frac{m \left|\mathbf{a}\right|}{V}.
\end{equation}
}

As an extension of the Bond number in \Cref{eq:Bond_number}, we define the confinement Bond number as $\mathrm{Bo} \mathrm{C}^2$. This combined dimensionless number effectively replaces the bubble length scale in the Bond number definition with the confinement ratio length scale. Additionally, we define the offset Bond number $\mathrm{Bo} \mathrm{B}_{\text{offset}}^2$, which has a similar purpose to the confinement Bond number. However, here the bubble length scale is replaced by the offset length scale between two bubbles.

\section{Analytical model}\label{sec:analytical-model}

\subsection{Critical Bond number}\label{sec:critical-Bond-number}
In this study, we focus on distinguishing whether bubbles pass the constriction. To this end, we define a critical Bond number, $\mathrm{Bo}_{\text{cr}}$, as the minimum value required for a single bubble to pass. For $\mathrm{Bo} < \mathrm{Bo}_{\text{cr}}$, a single bubble gets trapped upstream of the constriction; for $\mathrm{Bo} > \mathrm{Bo}_{\text{cr}}$, it passes.

We estimate the critical Bond number using a simple analytical model that relates the Laplace pressure due to bubble deformation induced by the constriction and gravity-induced driving forces.
Viscous contributions to the pressure are neglected since velocities are small when a single bubble is passing through the constriction with a Bond number close to the critical one.
Since bubble crossing occurs on time scales much longer than pressure equilibration across the constriction, we assume a uniform background pressure.
Accordingly, the \newtext{pressure balance on the bubble over the constriction inlet} reads
\begin{equation}
    \frac{F(z)}{A} = P_g - P_{L}(z),
    \label{eq:pressure-balance}
\end{equation}
where $z$ denotes the position of its center of mass. The gravity-induced pressure,
\begin{align}
    P_g = \frac{F_g}{A},
\end{align}
consists of the gravity-induced force on the bubble $F_g = m \left|\mathbf{a}\right|$, divided by an effective surface area $A$. \newtext{One might be trivially inclined to use the cross-sectional area of the constriction, $A_c = \frac{1}{4} \pi d^2$, as the effective surface area. However, this leads to incorrect limits for the gravity-induced pressure when $d$ approaches zero. To this end, we pick the effective surface area to be the cross-section of an undeformed bubble $A = \pi R^2$.}

\begin{figure}[!t]
    \centering
    \includegraphics{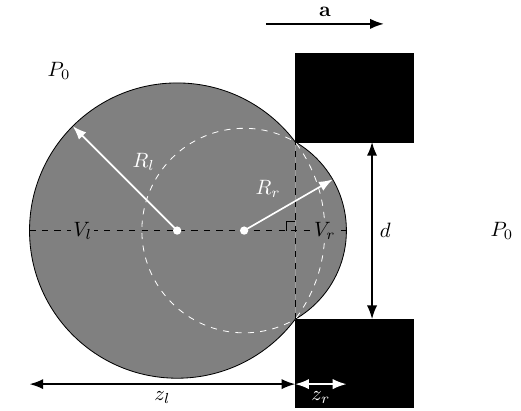}
    \caption{
    Two-dimensional schematic of a bubble entering a constriction. The left and right spherical caps have radii $R_l$ and $R_r$, with maximum distances from the channel inlet $z_l$ and $z_r$. The height of the constriction is $d$, and the cap volumes are $V_l$ and $V_r$. Finally, the background pressure is denoted by $P_0$, and the bubble experiences a gravity-induced acceleration $\mathbf{a}$.
    }
    \label{fig:schematic}
\end{figure}

\newtext{To compute the Laplace pressure contribution, we assume a uniform internal pressure and project the Laplace pressures on the cross-section of the constriction inlet. To this end, in contrast to the previously considered pressures, the Laplace pressures do act on the cross-sectional area of the constriction, $A_c$. We can write the Laplace pressure differential over the constriction inlet as:}
\begin{equation}
    P_{L}(z) = \frac{2\sigma}{R_r(z)} - \frac{2\sigma}{R_l(z)}.
\end{equation}
\newtext{Here,} $R_l(z)$ and $R_r(z)$ are the radii of the left and right spherical caps, respectively. This is shown schematically in \Cref{fig:schematic}, where the entry of the bubble into the constriction is illustrated.

Using \Cref{eq:pressure-balance}, we define the critical Bond number
\begin{equation}
    \mathrm{Bo}_{\text{cr}} = \frac{3}{2} \left(\frac{R}{R_r} - \frac{R}{R_l}\right)
    \label{eq:critical-Bond-number}
\end{equation}
as the value of $\mathrm{Bo}$ for which $F(z) = 0$.
In the following, we focus on small bubbles whose volume is conserved during translocations. Accordingly, this leads to a relation between the left and right spherical cap radii via volume conservation. Given that the maximum resistance pressure is obtained when $R_r = d / 2$, we obtain a volume
\begin{equation}
    V_r = \frac{1}{12} \pi d^3
    \label{eq:volume-right}
\end{equation}
for the right spherical cap. For the left spherical cap, we have Pythagoras' theorem to compute the maximum distance to the channel inlet
\begin{equation}
    z_l = R_l + \sqrt{R_l^2 - \left(\frac{d}{2}\right)^2},
    \label{eq:z-left}
\end{equation}
which can be rewritten as
\begin{equation}
    R_l = \frac{d^2 + 4 z_l^2}{8 z_l}.
    \label{eq:radius-left}
\end{equation}
Equating the volume of the left and right spherical caps with the total volume, we get
\begin{equation}
    \frac{1}{6} \pi z_l^3 + \frac{1}{8} d^2 \pi z_l + \frac{1}{12} d^3 \pi - \frac{4}{3} \pi R^3 = 0,
    \label{eq:z-left-cubic-equation}
\end{equation}
which can then be solved for $z_l$. We do this by making use of the Cardano method, inspired by a similar derivation for pressure-driven flow by Zhang \textit{et al.}~\cite{Zhang2018PressureViscousDroplet}, which gives
\begin{equation}
    z_l = \frac{p^2 - d^2}{2 p},
    \label{eq:z-left-alpha}
\end{equation}
with
\begin{equation}
    p = \left(32 R^3 + \sqrt{1024 R^6 - 128 d^3 R^3 + 5 d^6} - 2 d^3\right)^\frac{1}{3}.
    \label{eq:alpha}
\end{equation}
Performing a back substitution into \Cref{eq:radius-left} and \Cref{eq:critical-Bond-number} and with the definition of the confinement ratio, we obtain
\begin{equation}
    \mathrm{Bo}_{\text{cr}} = \frac{3\left(\mathrm{C}^2 + \mathrm{C} q - q^2\right)^2}{2 \mathrm{C} \left(\mathrm{C}^4 - \mathrm{C}^2 q^2 + q^4\right)}
    \label{eq:critical-Bond-number-explicit}
\end{equation}
as an expression for the critical Bond number with
\begin{equation}
    q = \left(4 + \sqrt{5 \mathrm{C}^6 - 16 \mathrm{C}^3 + 16} - 2 \mathrm{C}^3\right)^\frac{1}{3}.
    \label{eq:beta}
\end{equation}
\newtext{As can be seen, \Cref{eq:critical-Bond-number-explicit} depends solely on the confinement ratio $\mathrm{C}$. The part of this curve that is relevant for the simulations is plotted in \Cref{fig:state_diagram_d1}. In addition, interestingly enough \Cref{eq:critical-Bond-number-explicit} can be well approximated by 
\begin{align}
    \text{Bo}_\text{cr} \approx \frac{3}{2}\frac{1-\mathrm{C}}{\mathrm{C}},
    \label{eq:critical-Bond-number-approximated}
\end{align}
allowing the function to be interpreted more straightforwardly. In particular, \Cref{eq:critical-Bond-number-approximated} gives a similarly shaped curve and complies with the two limits that one would expect of \Cref{eq:critical-Bond-number-explicit}. When $\mathrm{C}$ approaches zero, i.e., the constriction diameter in \Cref{eq:confinement_ratio} becomes very small with respect to the bubble diameter, $\mathrm{Bo}_{\text{cr}}$ goes to infinity. On the other hand, when $\mathrm{C} = 1$ we obtain $\mathrm{Bo}_{\text{cr}} = 0$.}

As mentioned earlier, we expect \Cref{eq:critical-Bond-number-explicit} to hold as long as the viscous forces can be disregarded. Additionally, it should be noted that the assumption that the left and right caps of the bubble undergo perfectly spherical deformation is incorrect under certain conditions. In particular, when the magnitude of the driving force is similar to or larger than the magnitude of the surface tension forces, spherical deformation cannot be assumed. This will be further elaborated upon in \Cref{sec:single-bubble}, where simulation results are compared with the analytical predictions.

\subsection{Passage times}
For the cases where $\mathrm{Bo} > \mathrm{Bo}_{\text{cr}}$ the bubble will pass the constriction in a time $t_p$, which is defined as
\begin{align}
    t_p = \int_0^l \frac{\dd{z}}{v(z)},
\end{align}
where
\begin{align}
    v(z) = \frac{F(z)}{\gamma(z)}
\end{align}
is the local velocity and $\gamma(z)$ is the local friction coefficient. To keep the model simple and to derive estimations of the passage time, we assume the friction coefficient to be homogeneous $\gamma(z) = \gamma_0$. \newtext{While this assumption is likely too simplistic to obtain accurate predictions of the passage time, it keeps our analysis analytically tractable and allows for the identification of a master curve.} We approximate the force with its minimum value, which is attained at the entrance of the constriction. In fact, in this configuration, the opposing Laplace pressure at the front interface is at its maximum. In contrast, the Laplace contribution pushing from the back is at its minimum. Accordingly, we approximate the passage time as
\begin{align}
    t_p = \frac{l\gamma_0}{F_\text{min}},
    \label{eq:passage_time}
\end{align}
with $F_\text{min}$ being the smallest value of $F(z)$ in \Cref{eq:pressure-balance}. We can now substitute \Cref{eq:pressure-balance} into \Cref{eq:passage_time} to obtain
\begin{align}
   t_p = \frac{l\gamma_0}{A(P_g-P_{L})}.
   \label{eq:pressure-balance-passage_time}
\end{align}
Additionally, the contribution of the gravity-induced and Laplace pressures can be rewritten in terms of the Bond number as
\begin{equation}
    P_g - P_{L} = \left(\mathrm{Bo} - \mathrm{Bo}_{\text{cr}}\right) \frac{4}{3}\frac{\sigma}{R}.
    \label{eq:reference-pressure}
\end{equation}
Finally, by substituting \Cref{eq:reference-pressure} into \Cref{eq:pressure-balance-passage_time} and combining variables we retrieve
\begin{align}
    t_p = \frac{l\gamma_0}{\mathrm{Bo}-\mathrm{Bo}_\text{cr}} \frac{3}{4\pi \sigma}.
    \label{eq:t_zero}
\end{align}
For $\mathrm{Bo} \gg \mathrm{Bo}_\text{cr}$, \Cref{eq:t_zero} leads to
\begin{align}
    t_0 \simeq \frac{l\gamma_0}{\mathrm{Bo}} \frac{3}{4\pi \sigma},
\end{align}
as an expression for the nominal passage time with negligible holdup. Accordingly, we can define a dimensionless passage time as
\begin{align}
    \frac{t_p}{t_0} = \frac{\mathrm{Bo}}{\mathrm{Bo}-\mathrm{Bo}_\text{cr}}.
    \label{eq:dimensionless-passage-time}
\end{align}

\section{Numerical method}\label{sec:numerical-method}
\subsection{Multiphase lattice Boltzmann method}
To simulate bubbles moving through a constriction, we make use of the well-established lattice Boltzmann method for which numerous extensions for multiphase and multicomponent flows exist~\cite{Benzi1992LatticeBoltzmannEquation,Kruger2017LatticeBoltzmannMethod,Liu2016MultiphaseLatticeBoltzmann}. Different variations of the method were applied to droplets passing through capillaries with fractionally wet walls and surface roughness~\cite{Imani2022ThreedimensionalSimulationDroplet, Imani2023EffectRoughnessDroplet}, to single droplets squeezing through a constriction using phase-field formulations~\cite{Yuan2019DynamicBehaviorDroplet, Yi2023EffectPhysicalProperties,Guo2026MorphologyTransportDynamics}, or to study emulsions in constricted geometries using a pseudopotential model~\cite{Wei2020FlowBehaviorsEmulsions}. Furthermore, bubble formation and transport in catalytic materials~\cite{Scheel2024EnhancementBubbleTransport} or snap-off effects occurring for individual droplets were investigated using a color-gradient model~\cite{Gu2023NumericalStudyDroplet}.
Here, we also apply a color-gradient model to investigate bubble transport due to the model's inherent stability of interfaces~\cite{Gunstensen1992MicroscopicModelingImmiscible,Lishchuk2003LatticeBoltzmannAlgorithm,Leclaire2017GeneralizedThreedimensionalLattice}.

We consider two immiscible fluids, related to the bubbles $b$ and carrier fluid $c$, each evolving according to its own distribution functions $f_i^b$ and $f_i^c$. For fluid $k = b$ or $c$, mass density and momentum are computed from these distribution functions as
\begin{equation}
    \rho_k = \rho_0 \sum_i f_i^k \quad \text{and} \quad \rho_k \boldsymbol{u}_k = \rho_0 \sum_i f_i^k \boldsymbol{c}_i
\end{equation}
with $\rho_0 = 1$ being the unit mass density. Furthermore, the total mass density and total momentum are computed as
\begin{equation}
    \rho_\text{tot} = \sum_k \rho_k \quad \text{and} \quad \rho_\text{tot} \boldsymbol{u}_\text{tot} = \sum_k \rho_k \boldsymbol{u}_k.
\end{equation}
The distribution functions of both fluids evolve with the lattice Boltzmann equation
\begin{equation}
  f_i^k\left(\boldsymbol{x}+\boldsymbol{c}_i \Delta t, t+\Delta t\right)=f_i^k(\boldsymbol{x}, t)+\Omega_i^k,
\label{eq:lattice_boltzmann_equation}
\end{equation}
where $f_i^k$ is the single particle distribution function at position $\boldsymbol{x}$ and time $t$~\cite{Benzi1992LatticeBoltzmannEquation}. We make use of a D3Q19 lattice, where the weight of every lattice vector $\boldsymbol{c}_i$ is denoted by $w_i$, and the speed of sound is given by $c_s = \frac{1}{\sqrt{3}}\frac{\Delta x}{\Delta t}$~\cite{Qian1992LatticeBGKModels}. Additionally, we set the lattice constant and time step to unity for simplicity ($\Delta x = \Delta t = 1$). Furthermore, the collision operator $\Omega_i^k$ is split into a summation of three independent collision operators as
\begin{equation}
    \Omega_i^k = \left(\Omega_i^k\right)^{\left(1\right)} + \left(\Omega_i^k\right)^{\left(2\right)} + \left(\Omega_i^k\right)^{\left(3\right)},
    \label{eq:operator_decomposition}
\end{equation}
being the single-phase collision, perturbation, and recoloring operators, respectively. This results in a sequence of collision steps, as described below.

Firstly, we apply the single-phase collision operator according to the well-known BGK formalism \cite{Bhatnagar1954ModelCollisionProcesses}
\begin{equation}
    \left(\Omega_i^k\right)^{\left(1\right)} = -\frac{f_i-f_i^{\mathrm{eq}}\left(\rho_k, \boldsymbol{u}_\text{tot}\right)}{\tau},
    \label{eq:operator_single_phase_collision}
\end{equation}
in order to reproduce solutions of the Navier-Stokes equations. \newtext{The relaxation time $\tau$ is related to the kinematic viscosity as
\begin{equation}
    \nu = c_s^2 \left(\tau - \frac{\Delta t}{2}\right).
    \label{eq:kinematic_viscosity}
\end{equation}
In our simulations,} we set $\tau = 1.0$ \newtext{and $\Delta t = 1$}. \newtext{As a result,} we do not have to work with a color-blind distribution and can apply this collision operator to both sets of distribution functions separately. \newtext{It also fixes the kinematic viscosity, but the high surface tensions that we use ensure that the capillary number in \Cref{eq:capillary_number} remains much smaller than one.} The equilibrium distribution $f_i^{\mathrm{eq}}$ is computed from the standard second-order truncated equilibrium distribution function
\begin{equation}
    f_i^{\mathrm{eq}}\left(\rho, \boldsymbol{u}\right) = w_i \rho\left[1+\frac{\boldsymbol{u} \cdot \boldsymbol{c}_i}{c_s^2}+\frac{\left(\boldsymbol{u} \cdot \boldsymbol{c}_i\right)^2}{2 c_s^4}-\frac{\boldsymbol{u} \cdot \boldsymbol{u}}{2 c_s^2}\right]\,,
    \label{eq:equilibrium_distribution}
\end{equation}
where $\rho$ and $\boldsymbol{u}$ correspond to the according densities and velocities.

Secondly, the perturbation operator is applied, which is implemented as a body force by means of the exact difference scheme proposed by Kupershtokh \textit{et al.}~\cite{Kupershtokh2009EquationsStateLattice}
\begin{equation}
    \left(\Omega_i^k\right)^{\left(2\right)} = f_i^{\mathrm{eq}}\left(\rho_k, \boldsymbol{u}_\text{tot} + \Delta\boldsymbol{u}_k\right) - f_i^{\mathrm{eq}}\left(\rho_k, \boldsymbol{u}_\text{tot}\right).
    \label{eq:operator_perturbation}
\end{equation}
It generates an interfacial force in order to reproduce surface tension and serves to apply a gravity-induced acceleration $\mathbf{a}_{k}$. The exact difference scheme was chosen due to its indifferentiability properties. These make it possible to apply it to every distribution function separately and are essential in order not to violate consistent hydrodynamics \cite{Asinari2008ConsistentLatticeBoltzmann}. The velocity $\boldsymbol{u}_\text{tot}$ is computed from the total momentum of both fluid components and the velocity shift
\begin{equation}
    \Delta\boldsymbol{u}_k = \frac{\mathbf{F}_k}{\rho_k} \Delta t
    \label{eq:delta_u}
\end{equation}
is related to the force $\mathbf{F}_k$ applied on every fluid component separately. We define this force based on the continuum surface force concept as \cite{Brackbill1992ContinuumMethodModeling, Lishchuk2003LatticeBoltzmannAlgorithm}
\begin{equation}
    \mathbf{F}_{k} = \left(\frac{1}{2} \sigma H \nabla \rho^N\right) \frac{\rho_k}{\rho_\text{tot}} + \rho_k \mathbf{a}_{k}
    \label{eq:continuum_surface_force}
\end{equation}
Here, $H$ is the mean interface curvature between both fluids, which can be calculated as
\begin{equation}
    H=-[(\mathbf{I}-\mathbf{n} \otimes \mathbf{n}) \cdot \nabla] \cdot \mathbf{n},
    \label{eq:interface_curvature}
\end{equation}
with
\begin{equation}
    \mathbf{n} = - \frac{\nabla \rho^N}{\left|\nabla \rho^N\right|}
    \label{eq:color_gradient_normal}
\end{equation}
being the normalized gradient of the color function
\begin{equation}
    \rho^N\left(\boldsymbol{x}, t\right) = \frac{\rho_b\left(\boldsymbol{x}, t\right) - \rho_c\left(\boldsymbol{x}, t\right)}{\rho_b\left(\boldsymbol{x}, t\right) + \rho_c\left(\boldsymbol{x}, t\right)}.
\end{equation}
To approximate the partial derivatives in \Cref{eq:continuum_surface_force,eq:color_gradient_normal}, we make use of an isotropic finite difference scheme. Using the lattice structure, we compute the partial derivatives as
\begin{equation}
    \nabla \phi\left(\boldsymbol{x}, t\right) = \frac{1}{c_s^2} \sum_i w_i \boldsymbol{c}_i \phi\left(\boldsymbol{x}+\boldsymbol{c}_i \Delta t, t\right)
    \label{eq:finite_difference_scheme}
\end{equation}
for any of the required variables $\phi$.

Finally, immiscibility of the fluid components needs to be enforced. Although the perturbation operator imposes a surface tension between the two fluids, it does not guarantee immiscibility. Hence, it is required to apply additional recoloring operators \cite{Latva-Kokko2005DiffusionPropertiesGradientbased}
\begin{equation}
    \begin{aligned}
        \left(\Omega_i^b\right)^{(3)} &= +\beta \frac{\rho_r \rho_b}{\rho_{\text{tot}}} w_i \frac{\boldsymbol{c}_i \cdot \mathbf{n}}{\left|\boldsymbol{c}_i\right|},\\
        \left(\Omega_i^c\right)^{(3)} &= -\beta \frac{\rho_r \rho_b}{\rho_{\text{tot}}} w_i \frac{\boldsymbol{c}_i \cdot \mathbf{n}}{\left|\boldsymbol{c}_i\right|}.
    \end{aligned}
    \label{eq:operator_recoloring}
\end{equation}
In this equation, $\beta$ is a parameter controlling the thickness of the numerical interface. To minimize the spurious currents, a value of $\beta = 0.7$ is used in all simulations.

\subsection{Boundary conditions}
At the inlet and outlet of the channel, we apply periodic boundary conditions. Furthermore, no-slip conditions are applied at solid boundaries and modelled in the lattice Boltzmann method by means of the halfway bounce-back rule. To this end, the streaming step of the fluid populations is locally modified to
\begin{equation}
    f_i^k\left(\boldsymbol{x}, t+\Delta t\right) = f_{\bar{i}}^k(\boldsymbol{x}, t)
    \label{eq:bounce_back}
\end{equation}
with $\bar{i}$ denoting the index of a population moving into the opposite direction of $i$ and into a solid node. As a result, the boundary is located approximately halfway between the solid and the fluid nodes \cite{Kruger2017LatticeBoltzmannMethod}.

Additionally, we impose a preferential wetting on the carrier fluid at the solid boundaries by following the procedure outlined by Akai \textit{et al.}~\cite{Akai2018WettingBoundaryCondition}. However, given that all walls in our simulations are flat, we only use the six lattice vectors aligned with the Cartesian directions to select nodes at the fluid-solid boundary and to compute normals. Hereafter, we continue with the same lattice-weighted color value computation in the solid boundary as well as the same color-gradient realignment procedures.

\subsection{Bubble tracking}\label{sec:bubble-tracking}
After the densities and color fields are computed, we use the selection criterion $\rho^N\left(\boldsymbol{x}, t\right) > 0.0$ to select lattice nodes that belong to bubbles. To track the bubbles and compute their properties, we use the Hoshen-Kopelman algorithm \cite{Hoshen1976PercolationClusterDistribution, Frijters2015ParallelisedHoshenKopelman} to compute the sites that belong to an individual bubble. This enables us to compute various properties, such as when coalescence and breakup events occur, as well as properties of individual bubbles.

\section{Simulation results}\label{sec:simulation-results}
In the simulations, the diameter of the capillary is set to $D = 30$ lattice sites, and its length to $L = 192$ lattice sites. This leads to a computational domain of $32 \cross 32 \cross 192$ sites when a solid layer is included at the domain boundaries perpendicular to the flow direction. We set the length of the constriction to $l = 4$ lattice sites, while the constriction diameter is varied across simulations. Together with an initial bubble radius $R = 10$ lattice sites, this gives $\mathrm{C}_l = 0.2$, as discussed in \Cref{sec:problem-statement}.
Furthermore, we set an initial carrier fluid and bubble density $\rho_\text{init} = 1.0$, and apply a gravity-induced acceleration $\mathbf{a}_{b} = \left(\begin{array}{ccc} 0 & 0 & 5 \cdot 10^{-5} \end{array}\right)^{\top}$ per time step on the bubbles, mimicking a gravity-induced body force along the channel direction. Additionally, we vary the surface tension $\sigma$ between $0.05$ and $0.0005$, resulting in Bond numbers between $0.1$ and $10$. For all chosen Bond numbers, we set the confinement ratio from $0.3$ to $1.0$ in single-bubble simulations. For simulations with two bubbles, we also vary $\mathrm{B}_{\text{offset}}$ from $1.0$ to $3.0$ in steps of $0.5$. We run every simulation for $10^5$ time steps, which is found to be sufficient to predict the passage or clogging of single and paired bubbles.

\subsection{Single bubble}\label{sec:single-bubble}
\begin{figure}[!t]
    \centering
    \includegraphics{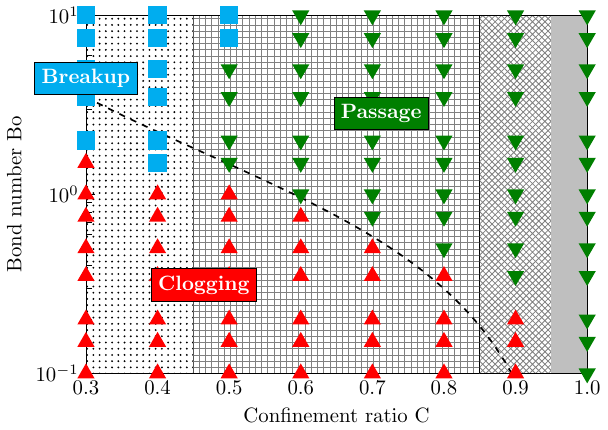}
    \caption{%The 
    State diagram of a single bubble passing through a constriction. \newtext{We plot the Bond number in \Cref{eq:Bond_number} against the confinement ratio in \Cref{eq:confinement_ratio} and} identify three different states: Red upward-facing triangles denote clogging of the channel, where the bubble is stuck in front of the constriction and is unable to enter. Green downward-facing triangles denote the passage of the bubble through the constriction. Blue squares correspond to bubble passage, with breakup into two smaller bubbles. Additionally, the gray background patterns indicate the four regions in the confinement ratio. From left to right, we have strong, moderate, weak, and no confinement. The dashed line is the analytical prediction for $\mathrm{Bo}_{\text{cr}}$ (\Cref{eq:critical-Bond-number-explicit}).}
    \label{fig:state_diagram_d1}
\end{figure}

We first simulate the passage of a single bubble through a circular constriction. \Cref{fig:state_diagram_d1} shows the state diagram of a single bubble passing through the capillaries. Three different states can be identified: \textit{clogging} of the constriction, \textit{passage} of the bubble, and \textit{breakup} of the bubble while passing the constriction.

\begin{figure}
    \centering
    \includegraphics{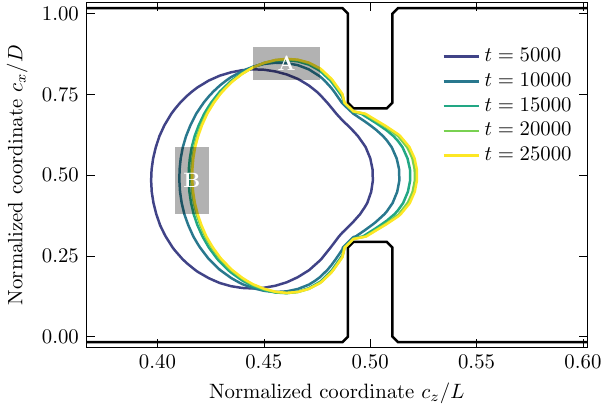}
    \caption{Single bubble deformation without passage: shown is the bubble profile for a central slice of the simulation box with confinement ratio $\mathrm{C} = 0.5$ and Bond number $\mathrm{Bo} = 1.0$. Only the region of the simulation box containing the bubble is shown, and the coordinates of the fluid nodes, $c_x$ and $c_z$, are normalized by the diameter $D$ and length $L$, respectively. The bubble approaches the constriction and clogs after around $25000 \Delta t$, with the left spherical cap deforming inhomogeneously to satisfy the imposed contact angle. Gray rectangles denote regions of increased (A) and reduced (B) curvature with respect to the spherical cap approximation.}
    \label{fig:contour_spherical_caps}
\end{figure}

\begin{figure}
    \centering
    \includegraphics{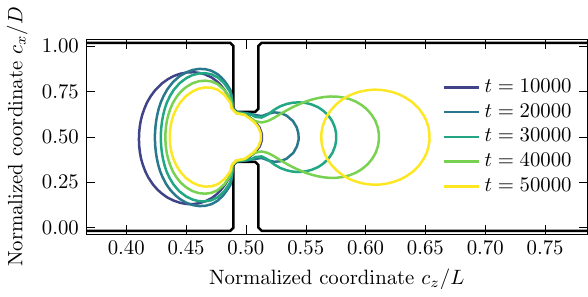}
    \caption{\newtext{Single bubble breakup: shown is the bubble profile for a central slice of the simulation box with confinement ratio $\mathrm{C} = 0.3$ and Bond number $\mathrm{Bo} = 2.0$. As for \Cref{fig:contour_spherical_caps}, we only show the region of the simulation box containing the bubble, and the coordinates of the fluid nodes, $c_x$ and $c_z$, are normalized by the diameter $D$ and length $L$, respectively. The bubble moves through the constriction, showing highly non-spherical deformation during passage. We observe bubble breakup between $40000 \Delta t$ and $50000 \Delta t$.}}
    \label{fig:contour_breakup}
\end{figure}

Moreover, four distinct regions can be identified based on the confinement ratio. Since our simulations are conducted with confinement ratio intervals of $0.1$, we distinguish: \textit{no confinement} ($\mathrm{C} \geq 1.0$), \textit{weak confinement} ($\mathrm{C} = 0.9$), \textit{moderate confinement} ($0.5 \leq \mathrm{C} \leq 0.8$), and \textit{strong confinement} ($\mathrm{C} < 0.5$). As expected, for simulations without confinement ($\mathrm{C} = 1.0$), the bubble passes the constriction independent of the Bond number. For simulations with confinement ($\mathrm{C} < 1.0$), we observe that the passage of the bubble through the constriction becomes Bond number dependent. However, when comparing the simulation results with \Cref{eq:critical-Bond-number-explicit}, we find that the equation underestimates the separation between the two states, clogging and passage, for weak confinement. This is likely due to the exclusion of viscous effects in its derivation. For moderate confinement, \Cref{eq:critical-Bond-number-explicit} provides accurate predictions of the separating line between clogging and passage. Even though we see some deviations from the assumption that the deformation of the spherical caps is perfectly spherical, as further illustrated in \Cref{fig:contour_spherical_caps}. For strong confinement, \Cref{fig:state_diagram_d1} shows that the bubble does not cross anymore. For these values of $\mathrm{C}$, only a separation between clogging and breakup can be observed, which comes with strong \newtext{non-spherical} deformations of the bubble. \newtext{This is illustrated in \Cref{fig:contour_breakup}, where we show a contour plot of bubble breakup during passage. A breakup occurs due to a major discrepancy in the resistance imposed by the constriction between the leading and trailing part of a bubble. Therefore, the leading part of a bubble experiences stronger accelerations than the trailing part. When this effect is strong enough, it can overcome surface tension and lead to interface thinning in the center. In that case, carrier fluid flows in inbetween the leading and trailing part, which can eventually lead to bubble breakup. Given the strong non-spherical deformations that are present in this case}, \Cref{eq:critical-Bond-number-explicit} is not expected to provide a reasonable prediction on the passage of the bubble in this region \newtext{of the state diagram}.

\begin{figure}[!t]
    \centering
    \begin{tabular}{cc}
        \begin{subfigure}[t]{.49\linewidth}
        \centering
        \includegraphics{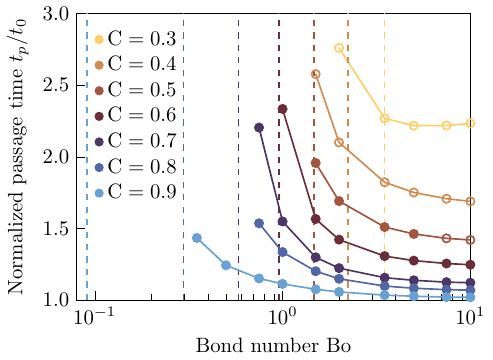}
        \caption{}
        \label{fig:passage_times_d1_asymptotes}
        \end{subfigure}%
        &
        \begin{subfigure}[t]{.49\linewidth}
        \centering
        \includegraphics{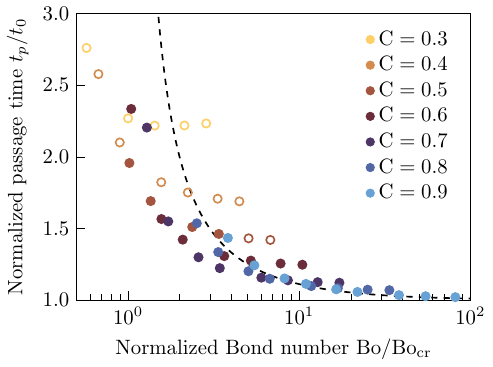}
        \caption{}
        \label{fig:passage_times_d1_curve}
        \end{subfigure}
    \end{tabular}
    \caption{Normalized passage time vs. Bond number \textbf{(a)} and normalized Bond number \textbf{(b)} for bubble passage with and without breakup. Open symbols denote passage with breakup, while the closed symbols denote passage without breakup. Colors denote different confinement ratios, where the dashed lines show the critical Bond number, obtained from \Cref{eq:critical-Bond-number-explicit}. \newtext{Three} reasons for deviations from the analytical predictions can be identified: viscous effects, violations of the spherical cap approximation\newtext{, and a non-constant friction coefficient.}}
    \label{fig:passage_times_d1}
\end{figure}

Regarding the passage time of the bubble through the constriction, we first investigate the asymptotic behavior as a function of the Bond number. In \Cref{fig:passage_times_d1_asymptotes}, the normalized passage time $t_p/t_0$ is plotted against the Bond number, $\mathrm{Bo}$. In the simulations, the bubble's passage time is computed by measuring the time required for the bubble's center of mass to move from a distance $R$ in front of the constriction to a distance $R$ beyond it. If the bubble breaks while passing through the constriction, we use the center of mass of the leading bubble. We then normalize this simulated passage time with the time a bubble takes to cross the same distance in a similar simulation with $\mathrm{C} = 1.0$. When $\mathrm{C} = 1.0$, the passage time of the bubble is governed by changes in Bond number, implemented by changes in surface tension, and viscous effects. Due to these changes in surface tension, the passage time slightly increases for smaller $\mathrm{Bo}$ when $\mathrm{C} = 1.0$. This is directly linked to a decrease in the ability of a bubble to deform while passing.

We observe that for $\mathrm{C} \geq 0.5$, the passage times diverge when the Bond number approaches its critical value predicted by \Cref{eq:critical-Bond-number-explicit}. However, in the case of strong confinement, and hence breakup, we see that the predicted asymptotic behavior is violated. As already discussed before, \Cref{eq:critical-Bond-number-explicit} does not cover this regime.

At last, in \Cref{fig:passage_times_d1_curve}, the normalized passage times for the bubble moving through the constriction are plotted against the normalized Bond number $\mathrm{Bo}/\mathrm{Bo}_{\text{cr}} < 10$. We compare the normalized passage times from the simulations with the analytical predictions from \Cref{eq:dimensionless-passage-time}. It can be seen that \Cref{eq:dimensionless-passage-time} works well for large normalized Bond numbers, but also tends to both underestimate and overestimate the passage time when this number gets closer to one. \newtext{Three} reasons can be identified for this observation. First, the analytics do not include any contribution of viscous effects. Since these are present in the simulations, this can lead to an underestimation of the passage time by \Cref{eq:dimensionless-passage-time}. Second, under certain conditions, the spherical cap approximation underlying the analytics is invalid. Depending on which cap deforms non-spherically, this can lead to both an underestimation and an overestimation of the passage time. \newtext{Third, our model for predicting the passage time is based upon the assumption that the friction coefficient remains constant when the bubble is passing the constriction. Although this is a strong simplification, it makes the model analytically tractable and allows for the identification of a master curve.}

Indeed, the left cap of the bubble sometimes needs to deform non-spherically to comply with the prescribed contact angle and to accommodate geometric pinning effects. These differences in curvature of the cap are further illustrated in \Cref{fig:contour_spherical_caps}, and lead to a decrease in Laplace pressure on the left side of the constriction. As a result, the passage time and the required Bond number for the bubble to pass are underestimated by the analytics. In addition, the right cap of the bubble can deform unevenly. This generally leads to a larger Laplace pressure contribution in \Cref{eq:pressure-balance}, decreasing the passage time in the simulations. Moreover, the right cap undergoes uneven deformation in simulations with bubble breakup, where non-spherical deformation is fundamental for bubble splitting.

\subsection{Bubble pairs}\label{sec:pairs-of-bubbles}

We now perform simulations analogous to those in \Cref{sec:single-bubble}, but with an additional bubble initialized behind the leading one. To quantify the impact of the second bubble, we redefine the state space in terms of deviations from the single-bubble case. On this basis, we identify five distinct states:

\begin{itemize}
    \item \textbf{Clogging:} A single bubble placed in front of the constriction results in clogging of the capillary. When a second bubble is placed after this bubble, both bubbles eventually coalesce, but the capillary remains clogged.
    \item \textbf{Passage:} A single bubble placed in front of the constriction passes the constriction, where we also include a possible breakup of the bubble while passing. When an additional bubble is placed behind the leading bubble, passage still occurs. We do not make a distinction regarding how the passage occurs; if there is a breakup along the way or both bubbles coalesce, we still put it under passage.
    \item \textbf{Coalescence-induced clogging:} A single bubble passes the constriction. However, when a second bubble is placed behind the initial bubble, the second bubble catches up and coalesces before the leading bubble can pass. The resulting larger bubble clogs the capillary.
    \item \textbf{Coalescence-induced unclogging:} A single bubble is unable to pass the constriction and clogs the capillary. When a second bubble is added, both bubbles coalesce, and the resulting larger bubble passes the constriction with or without breakup.
    \item \textbf{Hydrodynamic unclogging:} A single bubble blocks the constriction and is unable to pass. When a second bubble arrives, it pushes the leading bubble through without coalescing. This happens purely due to hydrodynamic effects. As the bubbles move closer, the pressure behind the leading bubble increases as fluid is squeezed out of the interbubble gap. When the leading bubble has passed, the same effect occurs in the other direction. The pressure in the interbubble gap decreases, which pulls the trailing bubble through.
\end{itemize}

Similar to our previous analysis for a single bubble, we also identify the same four different regions with respect to the confinement ratio. For the regions \textit{no confinement} and \textit{weak confinement}, changes in the state space are relatively straightforward and will be described in the paragraphs below. For \textit{moderate confinement} and \textit{strong confinement}, we found that the changes in the state diagram are more intricate. Hence, for these confinement ratios, we opt for plots to display the state space. Finally, we focus on changes in passage times observed in simulations of a single bubble versus those of pairs of bubbles.

As expected, placing a second bubble behind the first one has little influence on the outcomes in the \textit{no confinement} region. In all cases, when $\mathrm{B}_{\text{offset}} > 1.0$, both bubbles pass the constriction without breakup or coalescence. However, when the trailing bubble is initialized without an interbubble gap, $\mathrm{B}_{\text{offset}} = 1.0$, the bubbles instantly coalesce. This generally leads to the passage of the larger, combined bubble without breakup. However, in the particular case where $\mathrm{B}_{\text{offset}} = 1.0$ and $\mathrm{Bo} \leq 0.2$, we observe coalescence-induced clogging.

Switching to \textit{weak confinement} leads to a slightly different state space. First, we no longer encounter cases of coalescence-induced clogging. Instead, we find the passage state for $\mathrm{Bo} > 0.2$. This is in line with the state diagram for a single bubble in \Cref{fig:state_diagram_d1}. Nevertheless, even if the bubbles coalesce due to a sufficiently small $\mathrm{B}_{\text{offset}}$, they will still pass without breakup for $\mathrm{Bo} > 0.2$. On the contrary, for simulations at smaller Bond numbers, we find a separation between clogging and hydrodynamic unclogging. This separation appears to depend solely on the offset between the bubbles, with hydrodynamic unclogging occurring once $\mathrm{B}_{\text{offset}} \geq 2.0$. As a result, when $\mathrm{Bo} \leq 0.2$, bubbles might still be able to pass a capillary with a confinement ratio $\mathrm{C} = 0.9$, given that the spacing between the bubbles is large enough to inhibit coalescence.

\begin{figure}[!t]
    \centering
    \begin{tabular}{cc}
        \begin{subfigure}[t]{.49\linewidth}
        \centering
        \includegraphics{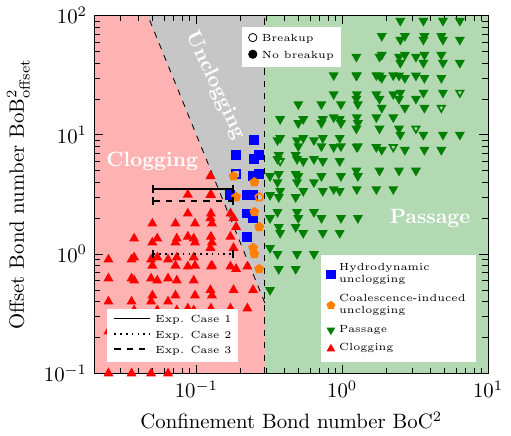}
        \caption{Moderate confinement ($0.5 \leq \mathrm{C} \leq 0.8$)}
        \label{fig:state_diagram_d2_moderate}
        \end{subfigure}%
        &
        \begin{subfigure}[t]{.49\linewidth}
        \centering
        \includegraphics{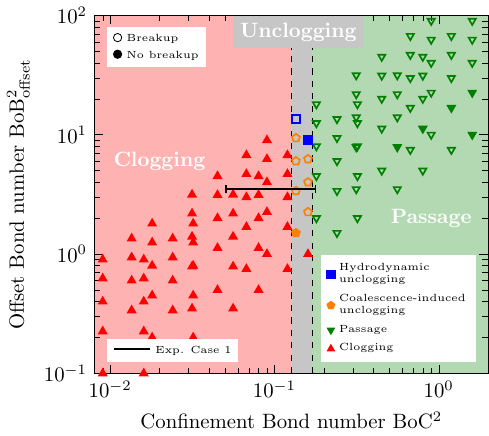}
        \caption{Strong confinement ($\mathrm{C} < 0.5$)}
        \label{fig:state_diagram_d2_strong}
        \end{subfigure}
    \end{tabular}
    \caption{State diagrams for two bubbles under \textbf{(a)} moderate confinement ($0.5 \leq \mathrm{C} \leq 0.8$) and \textbf{(b)} strong confinement ($\mathrm{C} < 0.5$). The marker shape indicates the state, while open or closed symbols distinguish between breakup and no breakup. 
    For moderate confinement, passage and clogging separate at $\mathrm{Bo}\mathrm{C}^2 = 0.295$. 
    Below this value, unclogging can occur if the offset Bond number exceeds the value predicted by \Cref{eq:power-law-separation}. For strong confinement, the transition from passage to clogging occurs at $\mathrm{Bo}\mathrm{C}^2 = 0.170$, with unclogging possible down to $\mathrm{Bo}\mathrm{C}^2 = 0.128$ if the offset Bond number exceeds one.
    For comparison, the range in the experimental Cases 1,\,2 and 3 (\Cref{tab:XR}) is indicated by the solid, dotted, and dashed horizontal lines, respectively.}
    \label{fig:state_diagram_d2}
\end{figure}

For \textit{moderate confinement}, the state space becomes too complex to describe directly. In \Cref{fig:state_diagram_d2_moderate}, the state space is plotted as a function of the offset Bond number and the confinement Bond number. Here, we switched to derived quantities of the Bond number in order to avoid the overlap of data points. We find a clean separation between passage and states with possible clogging for $\mathrm{Bo} \mathrm{C}^2 = 0.295$. This confinement Bond number is slightly lower than the maximum value of the critical confinement Bond number for moderate confinement, $\mathrm{Bo}_{\text{cr}} \mathrm{C}^2 = 0.369$ when $\mathrm{C} = 0.5$. As described in \Cref{sec:single-bubble}, this can again be explained due to a violation of the spherical cap assumption underlying this equation, and hence a reduction of the actual value of the critical Bond number.

For $\mathrm{Bo} \mathrm{C}^2 < 0.295$, single bubbles always clog the capillary for moderate confinement. However, if a second bubble moves in afterwards, there is a possibility of unclogging due to pressure build-up in the interbubble gap caused by the trailing bubble. We observe a correlation between the required offset Bond number and the confinement Bond number, and hence the probability for unclogging. Increasing the confinement Bond number increases the probability of unclogging. Moreover, increasing the offset Bond number increases the hydrodynamic pressure \newtext{that} a second bubble can build up behind the first, further boosting unclogging probability. A power law function, leading to the separating line in \Cref{fig:state_diagram_d2_moderate}, can be found between the clogging states and the unclogging states:
\begin{equation}
    \mathrm{Bo} \mathrm{B}_{\text{offset}}^2 = 10^{-2} \left(\mathrm{Bo} \mathrm{C}^2\right)^{-3}
    \label{eq:power-law-separation}
\end{equation}
This leads to a separation with only four outliers in the unclogging region. However, when unclogging occurs, the exact type is hard to predict based on offset and confinement Bond numbers alone. \newtext{As a result, the relation in \Cref{eq:power-law-separation} is purely empirical, describing the separation between clogging and unclogging without distinguishing between the responsible physical mechanisms. These are difficult to distinguish, given that they involve a complex interplay among surface-tension forces, bubble confinement, and interbubble lubrication forces. Nonetheless, \Cref{eq:power-law-separation} clearly shows that constrictions with smaller confinement ratios can be overcome when the distance between subsequent bubbles is greater. This is due to the trailing bubble catching up with larger momentum when the distance is greater. This increased momentum is transferred to the leading bubble through stronger interbubble lubrication forces, potentially supplemented by direct momentum exchange upon bubble coalescence. As evident from \Cref{fig:state_diagram_d2_moderate}, however, the confinement and offset Bond numbers alone do not allow us to predict whether unclogging is governed by purely hydrodynamic interactions or additionally involves bubble coalescence.} Additionally, it can be seen from \Cref{fig:state_diagram_d2_moderate} that \newtext{the} breakup of the bubble(s) does not play a significant role for moderate confinement. We observe breakups only for large Bond numbers or in very specific cases where both hydrodynamic unclogging and coalescence effects influence the dynamics simultaneously.

For \textit{strong confinement}, the corresponding state diagram is shown in \Cref{fig:state_diagram_d2_strong}. We observe that for a single bubble, the separation between passage and clogging can be directly predicted from the confinement Bond number, as is the case for moderate confinement. The separating line between these two states can be found at $\mathrm{Bo} \mathrm{C}^2 = 0.170$. As the confinement Bond number decreases, a single bubble always clogs the capillary. However, down to $\mathrm{Bo} \mathrm{C}^2 = 0.128$, there is a possibility for unclogging with pairs of bubbles. Besides, in contrast to moderate confinement, the offset Bond number can predict the type of unclogging that will occur. For offset Bond numbers between one and ten, we mainly observe coalescence-induced unclogging. However, as the offset Bond number approaches ten, hydrodynamic unclogging also becomes part of the state space. Subsequently, it should be noted that breakup is very prominent in cases where passage occurs. Passage without breakup is almost never observed for strong confinement.

\begin{figure}[!t]
    \centering
    \includegraphics{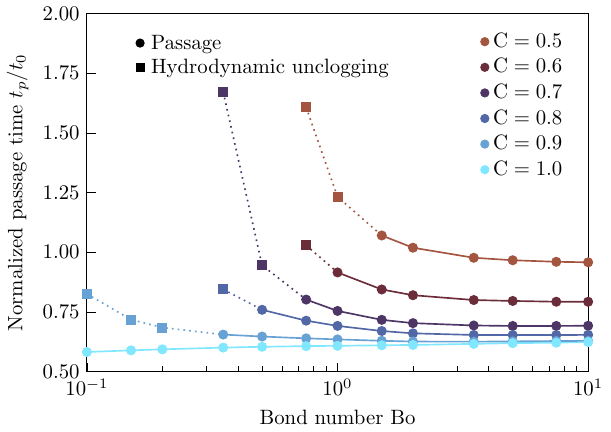}
    \caption{Normalized passage time of the leading bubble vs. Bond number. Passage times are normalized by the single-bubble passage time at confinement ratio $\mathrm{C}=1.0$ (as in \Cref{fig:passage_times_d1}).
    Results are shown for simulations without coalescence or breakup at $\mathrm{B}_{\text{offset}} = 3.0$. With negligible inertia, data for different offset ratios collapse onto a single curve. The change in linestyle marks the transition from passage to hydrodynamic unclogging, enabling passage for previously inaccessible combinations of Bond number and confinement ratio.
    }
    \label{fig:passage_times_d2}
\end{figure}

Finally, we investigate the passage time of the leading bubble when comparing pairs of bubbles with a single one. In \Cref{fig:passage_times_d2}, the normalized passage times of the leading bubble are plotted against the Bond number. The passage times are normalized with the passage time of a single bubble under no confinement, following the same procedures used in \Cref{fig:passage_times_d1}. To keep the passage of the leading bubble well-defined, we only consider cases where coalescence and breakup of bubbles do not occur, being simulations in the passage and hydrodynamic unclogging regimes. It can be seen that the passage of a bubble generally happens significantly faster when a trailing bubble is present. Besides, \Cref{fig:passage_times_d2} clearly shows that a trailing bubble enables bubble passage for combinations of Bond number and confinement ratio that are unreachable for a single bubble (hydrodynamic unclogging). Furthermore, it is important to note that the bubble offset ratio has a negligible influence on the observed passage time of the leading bubble. This further confirms that inertial effects do not play a significant role in our setup.

\section{Experiments}\label{sec:experiments}

Experimental investigations on the movement of bubbles through constricted capillaries are carried out as part of \mbox{X-ray} radiographic measurements of the gas transport in gas-liquid two-phase flows through open-porous structures. In particular, the model experiment addresses the motion of a chain of single gas bubbles through an open-porous nickel foam. At \SI{95}{\percent} open porosity, the estimated average pore diameter is \SI{2.3}{\milli\meter} (Ni-0610) or \SI{1.4}{\milli\meter} (Ni-1116). Both nickel foam samples were purchased from Recemat and functionalised with a hydrophilic coating (hexamethyldisiloxane, HMDSO, $<\SI{0.1}{\micro\meter}$ in thickness) by plasma-enhanced chemical vapour deposition~\cite{Heinrich2025SurfaceFunctionalizationAdditively}. For X-ray radiography, the nickel foam is placed in a measuring cell made of acrylic glass, which is highly transparent for X-rays. The cell has a rectangular cross-section: \SI{80}{\milli\meter} wide and \SI{5}{\milli\meter} deep, i.e., in the X-ray beam direction, matching the foam sample size of \qtyproduct{80x80x5}{\milli\meter}. A single nozzle, made of stainless steel and \SI{0.3}{\milli\meter} in outer diameter, is centered at the bottom of the cell. The vertical distance between the nozzle tip and the foam sample is approximately \SI{15}{\milli\meter}. The cell is filled with deionised water, and the liquid level is measured \SI{120}{\milli\meter} above the nozzle tip. In this way, the open-porous structure of the foam sample is completely submerged and filled with liquid. Bubble chains are generated by injecting compressed air through the nozzle at a constant gas flow rate using a mass flow controller. Variations of the gas flow rate in different measurement runs result in different bubble diameters $2R$ and detachment frequencies, which, in turn, yield the center-to-center distance $r_{21}$ of consecutive bubbles, as listed in \Cref{tab:XR}.

We perform X-ray radiography using a high-power X-ray tube (ISOVOLT 450M1/25-55, GE Sensing \& Inspection Technologies), operated at $\SI{200}{\kilo\volt}$ tube voltage and $\SI{22.5}{\milli\ampere}$ tube current, providing a divergent beam of polychromatic X-rays. The image acquisition system consists of a scintillation screen (SecureX~HB, Applied Scintillation Technologies), a mirror and lens arrangement (custom build, TSO Thalheim Spezialoptik), and a sCMOS camera (pco.edge~5.5, PCO). X-ray image sequences are acquired at \num{150} frames per second, \SI{3}{\milli\second} exposure time, and \SI{0.06}{\milli\meter} image pixel size.

\begin{figure}[!t]
    \centering
    \includegraphics[height=8.0cm]{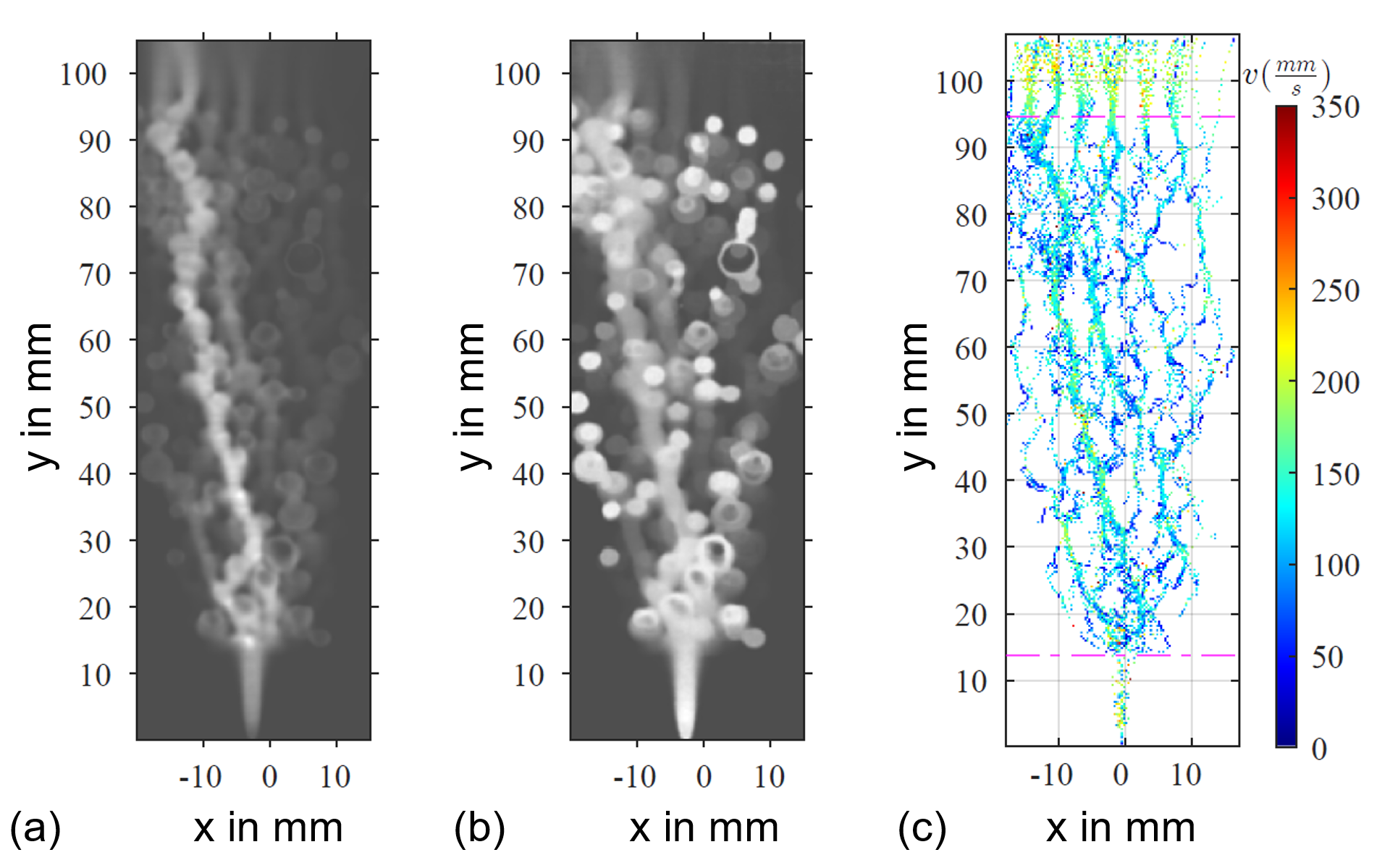}
    \caption{X-ray radiographic measurement of a chain of single bubbles through an open-porous nickel foam: time-averaged X-ray images highlighting either \textbf{(a)} the bubble motion paths or \textbf{(b)} immobile bubbles that are clogging pores; \textbf{(c)} local velocity of the bubbles along their paths.}
    \label{fig:XR}
\end{figure}

To measure the local velocity of individual bubbles as they move through the porous structure, the images are processed and analyzed in several steps. First, for every frame, the volumetric gas fraction in the measurement cell is mapped. This quantitative analysis of X-ray images, based on the measurement principle of transmission and attenuation of the beam intensity, is generally applicable to gas-liquid systems for quantifying phase fractions, as for example detailed in \cite{Skrypnik2025MeasurementLiquidFoam}. Second, we evaluate temporal changes in the gas fraction distribution across consecutive frames. This is the crucial step, acting as a filter to distinguish between bubbles that are moving through the porous structure (\Cref{fig:XR}a) and immobile bubbles that are clogging pores (\Cref{fig:XR}b). Third, individual bubbles can be tracked in consecutive frames, and the local velocity of each bubble is then calculated along its motion path (\Cref{fig:XR}c).

\begin{table}[!t]
    \centering
    \captionsetup{width=\textwidth}
    \caption{Experimental parameters and results. The constriction diameter $d$ is estimated based on the average pore diameter $D$. The average bubble diameter $2R$ and center-to-center distance $r_{21}$ are determined experimentally. This yields the confinement ratio $\mathrm{C}$, the bubble offset ratio $\mathrm{B}_{\text{offset}}$, the confinement Bond number $\mathrm{Bo} \mathrm{C}^2$ and the offset Bond number $\mathrm{Bo} \mathrm{B}_{\text{offset}}^2$. For each case, the comment on the bubble behaviour refers to the distinct states indicated in \Cref{fig:state_diagram_d2}, where the simulation results are shown in comparison with the range of the experimental data listed here.}
    \label{tab:XR}
    \begin{tabular}{ccccccccc}
        \hline
        Case & $D / \si{\milli\meter}$ & $d / \si{\milli\meter}$ & $2R / \si{\milli\meter}$ & $r_{21} / \si{\milli\meter}$ & $\mathrm{C}$ & $\mathrm{B}_{\text{offset}}$ & $\mathrm{Bo} \mathrm{C}^2$ & $\mathrm{Bo} \mathrm{B}_{\text{offset}}^2$ \\
        \hline
        1 & \num{2.3} & \numrange{1.16}{2.3} & \num{3.7} & \num{10.2} & \numrange{0.31}{0.62} & \num{2.8} & \numrange{0.05}{0.18} & \num{3.5} \\
         &  &  &  &  & \multicolumn{2}{l}{Strong to moderate confinement:} & \multicolumn{2}{l}{Clogging / Unclogging} \\
        2 & \num{2.3} & \numrange{1.16}{2.3} & \num{2.0} & \num{5.5} & \numrange{0.58}{1.15} & \num{2.7} & \numrange{0.05}{0.18} & \num{1.0} \\
         &  &  &  &  & \multicolumn{2}{l}{Moderate to no confinement:} & \multicolumn{2}{l}{Clogging / Unclogging} \\
        3 & \num{2.3} & \numrange{1.16}{2.3} & \num{1.9} & \num{9.1} & \numrange{0.61}{1.21}& \num{4.8} & \numrange{0.05}{0.18} & \num{2.8} \\
         &  &  &  &  & \multicolumn{2}{l}{Moderate to no confinement:} & \multicolumn{2}{l}{Clogging / Unclogging} \\
        4 & \num{1.4} & \numrange{0.71}{1.4} & \num{2.0} & \num{5.5} & \numrange{0.35}{0.70}& \num{2.7} & \numrange{0.02}{0.07} & \num{1.0} \\
         &  &  &  &  & \multicolumn{2}{l}{Strong to moderate confinement:} & \multicolumn{2}{l}{Clogging} \\
        5 & \num{1.4} & \numrange{0.71}{1.4} & \num{1.9} & \num{9.1} & \numrange{0.37}{0.74}& \num{4.8} & \numrange{0.02}{0.07} & \num{2.8} \\
         &  &  &  &  & \multicolumn{2}{l}{Strong to moderate confinement:} & \multicolumn{2}{l}{Clogging} \\
        \hline
    \end{tabular}
\end{table}

The experimental parameters and results are summarised in \Cref{tab:XR}, considering five different cases for comparison with the simulations. To this end, the averaged pore diameter of the nickel foam is referred to as $D$, similarly to the channel diameter defined in \Cref{fig:schematic_problem_statement}. We assume that the constriction diameter $d$ of individual pores is between $d_\text{min} = 0.51 \cdot D$ and $d_\text{max} = D$. The minimum $d_\text{min}$ is equivalent to the diagonal length of the square face of a Kelvin cell, as an example of an ordered monodisperse foam \cite{Cantat2013FoamsStructureDynamics}. This assumption includes that the volume of the Kelvin cell is equivalent to a sphere with the average pore diameter $D$. Furthermore, we calculate the confinement ratio $\mathrm{C}$ and the bubble offset ratio $\mathrm{B}_{\text{offset}}$ with the average bubble diameter $2R$ and the center-to-center distance $r_{21}$, as defined in \Cref{eq:confinement_ratio} and \Cref{eq:bubble_offset_ratio}. Proceeding from \Cref{eq:Bond_number}, the confinement Bond number %$\mathrm{Bo} \mathrm{C}^2 $
\begin{equation}
    \mathrm{Bo} \mathrm{C}^2 = \left( \frac{F_{g} R^{2}}{V\sigma} \right) \left( \frac{d}{2R} \right) ^2 = \frac{\rho_l \, g \, d^2}{4 \, \sigma} %\quad \text{,}
    \label{eq:Bo_C}
\end{equation}
and the bubble offset Bond number
\begin{equation}
    \mathrm{Bo} \mathrm{B}_{\text{offset}}^2 = \left( \frac{F_{g} R^{2}}{V\sigma} \right) \left( \frac{r_{21}}{2R} \right) ^2 = \frac{\rho_l \, g \, r_{21}^2}{4 \, \sigma} %\quad \text{,}
    \label{eq:Bo_B_offset}
\end{equation}
can be estimated with \mbox{$\rho_l = \SI{997}{\kilo\gram\per\meter\cubed}$}, \mbox{$\sigma = \SI{72}{\milli\newton\per\meter}$} and $g = \SI{9.81}{\meter\per\second\squared}$, %denoting 
referring to volumetric mass density, surface tension of deionised water, and gravitational acceleration. These two Bond numbers are then used in order to compare with the simulation results for moderate (\Cref{fig:state_diagram_d2_moderate}) and strong confinement (\Cref{fig:state_diagram_d2_strong}). The estimated constriction diameter and, thus, the confinement ratio naturally vary for all the experimental cases listed in \Cref{tab:XR} and discussed in the following. 

Case\,1 in \Cref{tab:XR} refers to the example shown in \Cref{fig:XR}. The average bubble diameter is larger than the average pore diameter, and the bubbles are always confined. At moderate confinement, here $\mathrm{C} = 0.62$, Case\,1 is within the state unclogging in \Cref{fig:state_diagram_d2_moderate}. This matches the experimental observation. The bubble tracking indicates that the unclogging is rather due to hydrodynamic effects than coalescence-induced. Furthermore, the \mbox{X-ray} images also point towards individual pores that are clogging, which is in line with the simulation results in the estimated ranges of $\mathrm{C}$ and $\mathrm{Bo} \mathrm{C}^2$.

In Cases\,2 and 3, the average bubble diameter is slightly smaller than the average pore diameter, but larger than the estimated minimum constriction diameter. The smaller bubbles are more likely to pass through the pore structure without confinement ($\mathrm{C} \geq 1.0$). However, individual pores temporarily clog and unclog, similar to Case\,1, which presumably is linked to weak to moderate confinement (\Cref{fig:state_diagram_d2_moderate}). This applies to both Case\,2 and 3, i.e., it is independent of the bubble offset ratio.

At a smaller average pore diameter (Case\,4 and 5), most of the bubbles get stuck, and the pores remain clogged, as predicted for $\mathrm{Bo} \mathrm{C}^2 < \num{0.1}$ at both moderate (\Cref{fig:state_diagram_d2_moderate}) and strong confinement (\Cref{fig:state_diagram_d2_strong}). Unclogging can only be observed after clustering and coalescence of multiple bubbles, thus forming an extraordinarily large bubble spanning several pore diameters. 

All in all, these experimental cases mainly point towards hydrodynamic unclogging, and bubble clustering suggests that coalescence-induced unclogging may be involved as well. Providing measurement-based experimental evidence for the coalescence of two consecutive bubbles in the porous structure requires an even higher spatial and temporal resolution than achieved in the X-ray image sequences analyzed here.

\section{Conclusion}\label{sec:conclusion}
We investigated the transport of gas bubbles through constricted geometries that serve as a simplified representation of pore throats in highly porous media such as porous transport layers (PTLs). 

Without confinement, bubbles generally pass, except at very low Bond numbers and small initial spacings, where coalescence-induced clogging can occur. Under weak confinement, viscous effects and increased Laplace pressure hinder passage. Although a single bubble may clog at low Bond number, sufficiently spaced pairs can pass via hydrodynamic unclogging. For moderate confinement, Laplace pressure dominates. Here, the confinement Bond number fully predicts single-bubble passage, while unclogging with an additional bubble can be described by a power-law, dependent on the offset Bond number and the confinement Bond number. Finally, under strong confinement, breakup typically occurs during passage. We find that the confinement Bond number remains the key predictor for the behavior of both single bubbles and pairs.

\mbox{X-ray} radiographic measurements of bubbles moving through an open-porous nickel foam align with the states predicted by the simulations, including clogging of individual pores followed by hydrodynamic unclogging. We also find strong evidence that bubbles traverse such structures in clusters via successive unclogging events. The coalescence-induced unclogging predicted by simulations is a potential reason behind this unclogging. However, a direct observation requires higher-resolution radiographic or tomographic studies.

Overall, this work identifies the key mechanisms governing bubble motion across pore-scale constrictions and establishes a dimensionless framework for predicting gas transport through highly porous media such as porous transport layers. These insights contribute to understanding two-phase transport limitations in electrochemical devices and may help guide the design of porous materials that promote efficient gas removal.

\begin{acknowledgments}
We acknowledge the Helmholtz Association of German Research Centers (HGF) and the Federal Ministry of Education and Research (BMFTR), Germany, for providing funding via the Innovation Pool project “Solar H2: Highly Pure and Compressed, and under Grant no. 03HY123E (H2Giga, SINEWAVE\,/\,OxySep). We also thank the Bavarian Ministry for Economy, State Development and Energy for funding of the project "H2Season” (RMF-SG20-3410-6-10-11). Additionally, we thank the Gauss Centre for Supercomputing e.V. (\url{www.gauss-centre.eu}) for providing computing time through the John von Neumann Institute for Computing (NIC) on the GCS Supercomputer JUWELS at J\"ulich Supercomputing Centre (JSC). \newtext{Finally, we want to thank Guanghao Jiao for his contribution to the experimental work.
The data presented in the figures of this paper is available on Zenodo \footnote{The data of the figures in this paper is available on Zenodo:~\href{http://dx.doi.org/10.5281/zenodo.18849902}{doi:10.5281/zenodo.18849902}}.}
\end{acknowledgments}

\bibliography{references}

\end{document}